\documentclass[aps,pra,nobalancelastpage,superscriptaddress,twocolumn]{revtex4-2}

\usepackage{color,ifthen,amsthm,amsmath,amsxtra,amsfonts,dsfont,graphicx,bm,tikz,scalerel,wasysym,physics,bbm,graphicx,amsthm,mathtools}
\usepackage[colorlinks=true,linkcolor=blue, citecolor=blue, urlcolor=blue, bookmarks]{hyperref}

\newcommand{\be}{\begin{equation}}
\newcommand{\ee}{\end{equation}}
\newcommand{\bea}{\begin{eqnarray}}
\newcommand{\eea}{\end{eqnarray}}
\newcommand{\bse}{\begin{subequations}}
\newcommand{\ese}{\end{subequations}}

\theoremstyle{plain}
\newcommand{\1}{\mathbbm{1}}

\usetikzlibrary{arrows,calc,matrix,backgrounds,fit,positioning,decorations.pathmorphing,decorations.markings,patterns,decorations.pathreplacing,shapes.misc}

\theoremstyle{plain}

\theoremstyle{plain}

\theoremstyle{plain}

\definecolor{half}{rgb}{0.95,0.95,0.95}
\definecolor{full}{rgb}{0,0,0}
\definecolor{halfborder}{rgb}{0.8,0.8,0.8}
\definecolor{border}{rgb}{0.3,0.3,0.3}
\definecolor{colUc}{rgb}{0.71,0.41,0.42}
\definecolor{colU}{rgb}{0.71,0.8,0.76}
\definecolor{FcolU}{rgb}{0.71,0.78,0.91}
\definecolor{colLines}{rgb}{0.31,0.31,0.31}
\definecolor{colMPSLines}{rgb}{0,0.01,0.18}
\definecolor{colMPS}{rgb}{0.27,0.45,0.77}
\definecolor{colIstLines}{rgb}{0.11,0.11,0.11}
\definecolor{colIstC}{rgb}{0.73,0.69,0.7}
\definecolor{colSt}{rgb}{0.96,0.74,0.59}
\definecolor{colIst}{rgb}{0.81,0.77,0.78}
\definecolor{colObs}{rgb}{1.,1.,1.}
\definecolor{colMPS}{RGB}{236,232,219}

\newcommand\tgridLine[4]{
    \draw [semithick,colLines] ({(#1)},{(#2)}) -- ({(#3)},{(#4)});
}

\newcommand\prop[3]{
    \draw[thick,colLines,fill=#3,rounded corners=1]
    ({((#1)-0.3)},{((#2)-0.3)}) rectangle ({((#1)+0.3)},{((#2)+0.3)});
}

\newcommand\ImpsM[3]{
    \draw[thick,colLines,fill=#3,rounded corners=0.5]
    ({((#1)-0.3)},{((#2)-0.15)}) rectangle ({((#1)+0.3)},{((#2)+0.15)});
}

\newcommand\Ips[3]{
    \draw[thick,colLines,fill=#3] ({(#1)},{(#2)}) circle (0.15);
}

\begin{document}

\title{Entanglement Negativity and Mutual Information after a Quantum Quench:\\
Exact Link from Space-Time Duality}
\date{\today}

\author{Bruno Bertini}
\affiliation{School of Physics and Astronomy, University of Nottingham, Nottingham, NG7 2RD, UK}
\affiliation{Centre for the Mathematics and Theoretical Physics of Quantum Non-Equilibrium Systems,
University of Nottingham, Nottingham, NG7 2RD, UK}

\author{Katja Klobas}
\affiliation{Rudolf Peierls Centre for Theoretical Physics, Clarendon Laboratory, Oxford OX1 3PU, UK}

\author{Tsung-Cheng Lu}
\affiliation{Perimeter Institute for Theoretical Physics, Waterloo, Ontario N2L 2Y5, Canada}

\begin{abstract}
We study the growth of entanglement between two adjacent regions in a tripartite, one-dimensional many-body system after a quantum quench. Combining a replica trick with a space-time duality transformation, we derive an exact, universal relation between the entanglement negativity and Renyi-1/2 mutual information which holds at times shorter than the sizes of all subsystems. Our proof is directly applicable to \emph{any} local quantum circuit, i.e., any lattice system in discrete time characterised by local interactions, irrespective of the nature of its dynamics. Our derivation indicates that such a relation can be directly extended to any system where information spreads with a finite maximal velocity. 
\end{abstract}

\maketitle

The Hilbert space of a quantum many-body system is frighteningly large --- its dimension scales exponentially with the volume --- therefore, it can be encoded in a classical computer only for very small system sizes. Fortunately, in most physically relevant systems the knowledge of the full Hilbert space is not necessary to describe equilibrium physics: one only needs a relatively small family of states characterised by low quantum entanglement~\cite{eisert2010colloquium}, which is efficiently described by \emph{tensor-network techniques}~\cite{schollwock2011density,cirac2021matrix}. 

The situation changes drastically for systems driven out of equilibrium. Even if the initial state admits an efficient description, quantum dynamics generically leads to growth of entanglement~\cite{calabrese2005evolution,liu2014entanglement,fagotti2008evolution,alba2017entanglement,alba2018entanglement,laeuchli2008spreading,kim2013ballistic,pal2018entangling,nahum2017quantum,chan2018solution,bertini2019entanglement,piroli2020exact,klobas2021entanglement}, which severely obstructs the simulability of systems at intermediate time scales. However, at sufficiently long times the systems are expected to achieve (generalised) thermalisation~\cite{polkovnikov2011,dalessio2016quantum,gogolin2016equilibration,eisert2015quantum,calabrese2016introduction,essler2016quench,vidmar2016generalized}, meaning that the reduced density matrix of a subregion becomes equivalent to an equilibrium ensemble. 
The latter has small mixed-state entanglement and admits an efficient description in terms of a matrix product operator (MPO)~\cite{Cirac_2004_mpo,Vidal_2004_mpo}. 
One therefore expects the amount of resources required to encode a time-evolving quantum many-body state to grow at early times, and decay at long times. This phenomenon has been described as \emph{entanglement
barrier}~\cite{dubail2017entanglement,wang2019barrier}, and has attracted
substantial attention as part of the ongoing effort to simulate quantum
dynamics for intermediate
times~\cite{prosen2007is,haegeman2011time,haegeman2016unifying,leviatan2017quantum,kloss2018time,white2018quantum,znidaric2019nonequilibrium,krumnow2019overcoming,rakovszky2020dissipationassisted,schmitt2022observations}.
Besides quantifying computational costs, the entanglement barrier also characterises the nature of the dynamics. For instance, its shape is qualitatively different for integrable and chaotic systems~\cite{wang2019barrier,reid2021entanglement}.

Due to its relevance for numerical simulations, the entanglement barrier has
been typically probed by means of the \emph{operator space
entanglement entropy} (OSEE)~\cite{prosen2007operator,pizorn2009operator}. Loosely speaking, the OSEE estimates the numerical cost to achieve a faithful MPO representation of the reduced density matrix~\cite{prosen2007operator,dubail2017entanglement,pizorn2009operator} but is not a measure of mixed-state entanglement. For instance, separable mixed states can have non-zero OSEE~\cite{prosen2007operator,dubail2017entanglement}.

In this Letter we adopt a more general quantum-information-theoretical point of
view and study the entanglement barrier through the lens of genuine mixed-state entanglement. Namely, we characterise the  entanglement dynamics by means of the entanglement negativity~\cite{peres1996separability,eisert1999comparison,horodecki2001separability,vidal2002computable,plenio2005logarithmic}, which has been used to explore several universal aspects of many-body physics~\cite{calabrese2012entanglement,Calabrese_2013_nega_extended,Alba_2013_cft_mc,Alba_2014_mc,calabrese2014finite,Calabrese_2016_negativity_spectrum, lee2013entanglement,wen2016edge,wen2016topological,hart2018entanglement,lu2020detecting,2021_Mulligan_negativity,lu_2022_topo_negativity, coser2014entanglement,eisler_2014_nega_dynamics,ryu_2015_nega_quench,Doyon_2015_nega_cft,ryu_2020_nega_contour,alba2019quantum,huse_2019_negativity,gruber2020time,lu2020entanglement,shi2020entanglement,kudler2020quasi,ryu_2020_correlation,murciano2021quench,sang2021entanglement,sharma2022measurement,weinstein2022measurement}.


In particular, we study the dynamics of the negativity between the regions $A$ and $B$ under a tripartition $ABC$ in \emph{generic} one-dimensional systems with discrete space-time and local interactions, i.e., local quantum circuits. Combining a replica trick~\cite{calabrese2012entanglement,calabrese2014finite} with a space-time duality approach~\cite{banuls2009matrix,muellerhermes2012tensor,hastings2015connecting,frias2022light,Akila_duality_2016,bertini2018exact,bertini2019entanglement,bertini2019exact,piroli2020exact,klobas2021entanglement,Chalker_2021_duality,garratt2021local,garratt2021manybody,Khemani_2021_duality,Lu_2021_duality,Khemani_2022_duality}, we provide a simple expression for the negativity, which is applicable for times smaller than the sizes of the regions $A$, $B$, $C$. We then use this expression to show that, up to exponentially small corrections, the negativity coincides with the R\'enyi-1/2 mutual information divided by two. This result holds for \emph{any} local quantum circuit, irrespective of the nature of the dynamics. Our findings generalise those of
Ref.~\cite{alba2019quantum} (see also
\cite{gruber2020time,lu2020entanglement,murciano2021quench}) --- where the aforementioned relation was discovered in non-interacting systems, and argued to hold
for all integrable systems --- and Refs.~\cite{ryu_2020_correlation, kudler2020quasi} --- where it has been observed in conformal field theories.

\begin{figure}
    \includegraphics[width=\columnwidth]{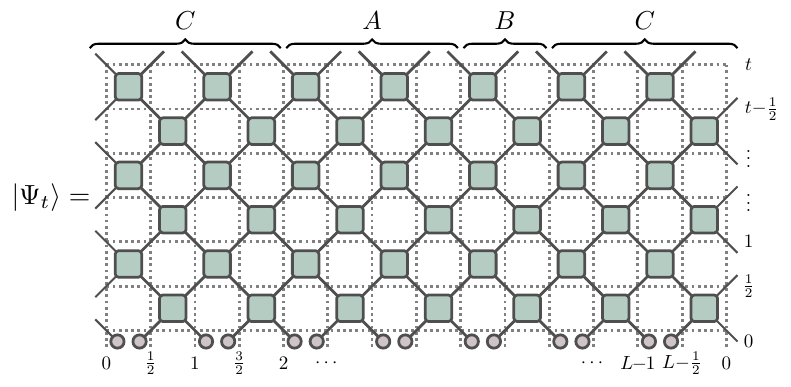}
    \caption{\label{fig:QC} Illustration of the time-evolution of a product state
    in a brick-work quantum circuit. Each time-step consists of the application of two-site
    gates first to odd and then to even pairs of consecutive sites, so that
    at time $t$ the state is given by $\ket{\Psi_t}=\mathbb{U}^{t} \ket{\Psi_0}$ (cf. Eq.~\eqref{eq:Floquet}). The boundary conditions are assumed to be \emph{periodic}.
    } 
\end{figure}

More specifically, we consider a quantum quench in a local quantum circuit of $2L$ qudits (see Fig.~\ref{fig:QC}), namely, a system where both space and time are discrete, each local site (labelled by integer and half-odd-integer numbers) hosts a $d$-state qudit, and there is a strict maximal speed $v_{\rm max}$ for the propagation of signals. 

Even though our argument does not depend on the specific
implementation of the circuit, for the sake of clarity we consider a brick-work quantum circuit where the time-evolution
operator $\mathbb U$ is written as 
\be  
\mathbb U = 
\smashoperator{\prod_{x \in \mathbb Z_{L}}} \eta_{x}(U) 
\smashoperator{\prod_{x \in \mathbb Z_{L}+\frac{1}{2}}}\eta_{x}(U) 
\,. 
\label{eq:Floquet}
\ee
Here $U$ denotes the \emph{local gate}, i.e., a $d^2\times d^2$ unitary matrix specifying the interactions between two neighbouring qudits, and the \emph{positioning operator} $\eta_{x}(\cdot)$ is a linear map that places a generic local operator $O$ on a periodic chain of $2L$ qudits such that its right edge is at position $x$. Our conventions imply $v_{\rm max}=1$.  In general, the local gate $U$ can be different at each space-time point describing systems with spatial disorder, or generic aperiodic driving. For the sake of clarity, however, in the main text we will assume $U$ to be homogeneous in the space-time, and take the initial state to be translationally invariant, $\ket{\Psi_0}=\ket{\psi}^{\otimes 2n}$. In the Supplemental Material (SM)~\footnote{See the Supplemental Material} we show how to extend our treatment to the inhomogeneous systems and matrix-product initial states.

In the following, we consider two adjacent,
finite, and non-complementary regions $A$ and $B$ in a tripartite system $ABC$ (see Fig.~\ref{fig:QC}), and characterise the entanglement between $A$ and $B$ via 
the \emph{logarithmic negativity}~\cite{peres1996separability,eisert1999comparison,horodecki2001separability,vidal2002computable}
\be
\mathcal{E}(t): =\ln {\rm tr}\sqrt{
    \left(\rho(t)_{AB}^{t_A}\right)^{\dagger}
    \rho(t)_{AB}^{t_A}
},
\label{eq:negativity}
\ee
where $\rho(t)_{AB}=\tr_{C}\ketbra{\Psi_t}{\Psi_t}$
is the reduced density matrix of the subsystem $A\cup B$ at time $t$ and $(\cdot)^{t_A}$ represents the partial transpose with respect to $A$. 

Our first objective is to obtain a more convenient expression for Eq.~\eqref{eq:negativity}. We proceed in two steps: First, we employ the replica trick by considering the even moments of the partially transposed density matrix \cite{calabrese2012entanglement,calabrese2014finite}
\be
\mathcal{E}_{2n}(t):=\ln {\rm tr}[(\rho(t)_{AB}^{t_A})^{2n}]
\label{eq:replicas}
\ee
for any positive integer $n$, which can then be reduced to the logarithmic negativity via an analytic continuation: $2n\mapsto \alpha$ followed by the limit $\alpha\to 1$. 

\begin{figure}
    \includegraphics[width=\columnwidth]{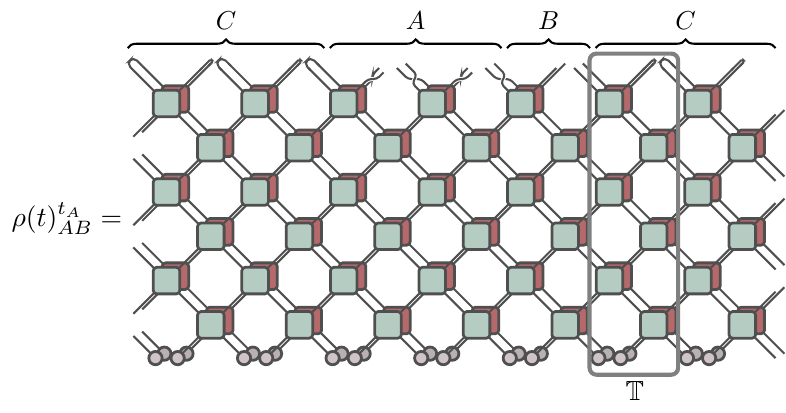}
    \caption{\label{fig:foldedCircuit} Folded representation of $\rho(t)_{AB}^{t_A}$.
    To obtain this diagram, we start with a copy of the state $\ket{\Psi_t}$ (green,
    cf.\ Fig.~\ref{fig:QC}), and its Hermitian conjugate $\bra{\Psi_t}$ (red), which is flipped and positioned behind $\ket{\Psi_t}$, i.e., the diagram is \emph{folded}.
    The degrees of freedom in the subsystem $C$ are then traced out, which is represented by connecting the legs of the two
    copies. Finally, we perform the partial transpose $(\cdot)^t_A$, by
    exchanging the output legs of the two copies in the subsystem $A$.
    }
\end{figure}

Second, we note that the quantities defined in  Eq.~\eqref{eq:replicas} can be conveniently computed using the space-time duality approach for the entanglement dynamics of Ref.~\cite{bertini2019entanglement} (see also~\cite{piroli2020exact,klobas2021entanglement}), see Fig.~\ref{fig:foldedCircuit} for a pictorial representation. 
\begin{figure*}
    \includegraphics[width=2\columnwidth]{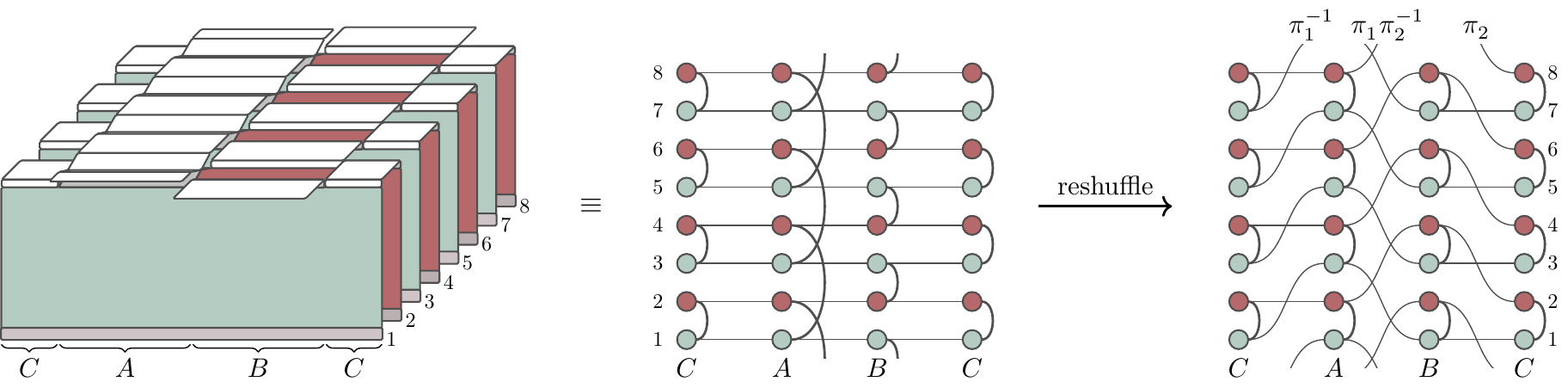}
    \caption{\label{fig:replicas} Schematic illustration of
    $\exp\left[\mathcal{E}_{2n}(t)\right]=\tr[(\rho(t)_{AB}^{t_{A}})^{2n}]$.
    In the left panel the green and red rectangles represent the time-evolved state and its conjugate respectively (i.e., the objects in Fig.~\ref{fig:foldedCircuit}), and the connections among different copies are depicted in white.  For clarity we introduce a simplified representation in the middle panel, where each copy is now represented by a row, and each column corresponds to one of the subsystems (right-most and left-most columns coincide due to periodic boundary conditions). Connections are represented by thick black lines, while the thinner horizontal lines connect different subsystems within the same copy. In the rightmost panel we reshuffled the positions of different copies so that the connections always occur between nearest neighbours. This is achieved by performing a permutation of copies at each interface between different subsystems (cf.\ Eq.~\eqref{eq:replicasrewritten}).}
\end{figure*}
We begin by introducing the \emph{space transfer matrix}
\be
\mathbb{T}= 
(\tilde{\mathbb{U}}\otimes \tilde{\mathbb{U}}^*)
\cdot
\mathbb{O}
\in {\rm End}(\mathcal{H}_t^{\otimes 2}),
\ee
where ${\rm End}(\mathcal{V})$ is the vector space of linear operators on the space $\mathcal{V}$, $(\cdot)^*$ represents complex conjugation, $\mathcal{H}_t=\mathbb{C}^{d^{2t+1}}$ denotes the Hilbert space of $2t+1$ qudits, $\mathbb{O}$ denotes the operator coupling the two copies of time-evolution, 
\be
\mathbb O :=\sum_{r,s=1}^d \tilde{\eta}_{t}(\ket{r}\!\!\bra{s})\!\otimes\! 
\tilde{\eta}_{t}(\ket{r}\!\!\bra{s}),
\ee
and $\tilde{\mathbb{U}}$ describes the evolution of one copy in the space direction,
\be
\tilde{\mathbb{U}} := 
\tilde{\eta}_0\left(\ketbra{\psi}{\psi}\right)
\mkern-8mu\smashoperator[r]{\prod_{\tau\in\mathbb{Z}_t+1\vphantom{\frac{1}{2}}}} 
\tilde{\eta}_\tau(\tilde{U})\mkern-6mu
\mkern-4mu\smashoperator[r]{\prod_{\tau\in\mathbb{Z}_t+\frac{1}{2}}}
\tilde{\eta}_{\tau}(\tilde{U}).
\ee
Here $\tilde U$ is obtained from the local unitary gate $U$ by applying the spacetime-duality transformation $[\tilde O]_{kl}^{ij} = O_{ki}^{lj}$, while $\tilde{\eta}_{\tau}(\cdot)$ denotes the ``dual'' positioning operator which places a local operator in $\mathcal{H}_t^2$, so that the \emph{top} edge is at position $\tau$ (see Fig.~\ref{fig:QC}
). A direct interpretation of the space transfer matrix $\mathbb{T}$ can be seen from Fig.~\ref{fig:foldedCircuit}: it implements the Heisenberg evolution when one exchanges the roles of space and time. Importantly, due to the unitarity of the time evolution, $\mathbb{T}$ has a unique maximal eigenvalue equal to one~\cite{muellerhermes2012tensor}.

Referring now to Fig.~\ref{fig:replicas}, Eq.~\eqref{eq:replicas} can be
expressed in terms of the space transfer matrix $\mathbb{T}$ as 
\be
\mathcal E_{2n} = \ln{\rm tr}[(\mathbb{T}_{2n}^{(\pi_1)})^{L_A}(\mathbb{T}_{2n}^{(\pi_2)})^{L_B}(\mathbb{T}_{2n})^{L_C}],
\label{eq:replicasrewritten}
\ee
where $L_S$ denotes the number of sites in $S$ divided by two and we introduced the operators  
\begin{align}
    &\mathbb{T}_{m} = \mathbb{T}^{\otimes m}, & &\mathbb{T}_{m}^{(\sigma)}= \mathbb P^{\dag}_{\sigma} \mathbb{T}_{m} \mathbb P_{\sigma}^{\phantom{\dag}},
\label{eq:transfermatrices}
\end{align}
and the permutations 
\be
\begin{aligned}
    \pi_1&=\begin{pmatrix}
1 & 2 & 3 & 4 & \cdots & 2m-1 & 2 m \\
2m-1 & 2 & 1 & 4  & \cdots & 2m-3 & 2 m\\
\end{pmatrix},\\
    \pi_2&=\begin{pmatrix}
1 & 2 & 3 & 4 & \cdots & 2m-1 & 2 m \\
1 & 2m & 3 & 2  & \cdots & 2m-1 & 2 m-2\\
\end{pmatrix}.
\end{aligned}
\ee
Here, $\mathbb P_{\sigma}$ is a unitary operator that acts on the multi-replica space $\mathcal{H}_t^{2m}$ and reshuffles copies according to
the permutation $\sigma$.  Namely, its action is given by
\be
\mathbb P_{\sigma} \ket*{{\boldsymbol s}_1}\otimes 
\cdots \otimes \ket*{{\boldsymbol s}_{2m}} = 
\ket*{{\boldsymbol s}_{\sigma(1)}}\otimes 
\cdots \otimes \ket*{{\boldsymbol s}_{\sigma(2m)}},
\ee
where $\ket{{\boldsymbol s}_j}$ are basis states of $\mathcal{H}_t$.
As pictured in Fig.~\ref{fig:replicas} the rewriting in Eq.~\eqref{eq:replicasrewritten}
essentially amounts to reordering the space-time sheets so that two copies connected by a contraction at the top are next to each other.
Importantly, this can be done only when the state $\ket{\Psi_0}$ is \emph{pure}, otherwise space-time sheets are connected also to different copies at the bottom. In fact, this is the \emph{only} ingredient needed for the validity of Eq.~\eqref{eq:replicasrewritten}. No other property of the dynamics or the initial state has been used. 

We now use the local structure of the dynamics to further simplify Eq.~\eqref{eq:replicasrewritten} in the early-time regime, i.e., when $v_{\rm max}t$ is smaller than the sizes of all subsystems $A, B, C$. Indeed, the existence of a maximal speed implies (see, e.g.,~\cite{klobas2021exactrelaxation,Note1})
\be
    \mathbb{T}^x=\ketbra{\rm r}{\rm l}, \qquad\qquad x\geq 2 v_{\rm max} t.
\label{eq:fixedpointrelation}
\ee
Here $\ket{\rm r}$ and $\bra{\rm l}$ denote the \emph{fixed points} of $\mathbb{T}$, i.e., the right/left eigenvectors corresponding to the eigenvalue one, and they are normalised such that $\braket{\rm l}{\rm r}=1$. 

From the physical point of view Eq.~\eqref{eq:fixedpointrelation} states that there are no correlations among regions out of the causal light cone. In the context of quantum many-body dynamics, Eq.~\eqref{eq:fixedpointrelation} has first been utilised in Ref.~\cite{banuls2009matrix} (see also \cite{muellerhermes2012tensor,hastings2015connecting}) to develop a numerical algorithm to describe the dynamics of local observables for large system sizes. Recently, it gained renewed attention due to a number of interesting developments: Refs.~\cite{bertini2019entanglement,piroli2020exact,klobas2021exact,klobas2021exactrelaxation,giudice2022temporal}
showed that the fixed points can be determined exactly in certain non-trivial examples including both integrable quantum circuits, such as quantum cellular
automaton Rule 54~\cite{bobenko1993two,buca2021rule}, and quantum chaotic ones, such as dual-unitary circuits~\cite{bertini2019exact}; Refs.~\cite{bertini2019entanglement,klobas2021entanglement, bertini2022growth} showed that the fixed points can be used to compute the slope of R\'enyi entropies following
quantum quenches; Refs.~\cite{lerose2021influence,lerose2021scaling,sonner2021influence,sonner2022characterizing} proposed a direct
physical interpretation of the fixed points based on the 
Feynman-Vernon influence functional.

In our context, Eq.~\eqref{eq:fixedpointrelation} implies that for 
\be
L_A,L_B,L_C\geq 2 v_{\rm max} t, 
\label{eq:relevantregime}
\ee
Eq.~\eqref{eq:replicasrewritten} reduces to   
\be
\!\!\!
\mathcal E_{2n}(t) = \ln\!\left[\!\!
\mel{{\rm l}_{2n}}{\mathbb P^{\dag}_{\!\pi_1}\!}{{\rm r}_{2n}}
\!\!
\mel{{\rm l}_{2n}}{\mathbb P^{\phantom{\dag}}_{\!\pi_1}\!\mathbb P^{\dag}_{\!\pi_2}\!}{{\rm r}_{2n}}
\!\!
\mel{{\rm l}_{2n}}{\mathbb P^{\phantom{\dag}}_{\!\pi_2}\!}{{\rm r}_{2n}}\!\right]\!, 
\label{eq:replicassimple}
\ee
where we defined $\ket{{\rm h}_m}= \ket{\rm h}^{\otimes m}$ with ${\rm h}={\rm r},{\rm l}$. 

To bring Eq.~\eqref{eq:replicassimple} to a form that can be analytically continued, 
we note that the states in $\mathcal{H}_t^{\otimes 2}$ can be viewed as matrices
in ${\rm End}(\mathcal{H}_t)$.
Formally this is achieved by a
vector-to-operator mapping defined on basis elements as (see, e.g., Ref.~\cite{bertini2018exact}) $\ket{\boldsymbol s}\otimes\ket{\boldsymbol r}\longmapsto \ketbra{\boldsymbol s}{\boldsymbol r}$. In particular, we introduce the matrices $M_l$ and $M_r$ with matrix elements
\begin{align}
 \mel{\boldsymbol s}{M_{{\rm h}}}{\boldsymbol r}=(\bra{\boldsymbol s}\otimes\bra{\boldsymbol r})\ket{\rm h}\,,\qquad {\rm h}={\rm l},{\rm r}. 
\label{eq:vtomapping}
\end{align}
Using Eq.~\eqref{eq:vtomapping} to rewrite the matrix elements in Eq.~\eqref{eq:replicassimple} we find 
\be
\begin{aligned}
\mel{{\rm l}_{2n}}{\mathbb P^{\dag}_{\!\pi_1}\!}{{\rm r}_{2n}} & 
    = \mel{{\rm l}_{2n}}{\mathbb P^{\phantom{\dag}}_{\!\pi_2}\!}{{\rm r}_{2n}} 
    =  {\rm tr}[(M_{\rm l}^\dag M^{\phantom{\dag}}_{\rm r})^{2n}],\\
\mel{{\rm l}_{2n}}{\mathbb P^{\phantom{\dag}}_{\!\pi_1}\!\mathbb P^{\dag}_{\!\pi_2}\!}{{\rm r}_{2n}} 
    & =  {\rm tr}[(M^\dag_{\rm l} M^{\phantom{\dag}}_{\rm r})^{n}]^2.
\end{aligned}
\ee
Therefore we arrive at the expression 
\be\label{eq:nega_moment}
\!{\mathcal E_{2n}}(t) = 2 \ln({\rm tr}[(M_{\rm l}^\dag M^{\phantom{\dag}}_{\rm r})^{2n}]{\rm tr}[(M_{\rm l}^\dag M^{\phantom{\dag}}_{\rm r})^{n}]),
\ee
valid in the regime given by Eq.~\eqref{eq:relevantregime}. Now performing the analytic continuation $\mathcal E_{2n}(t)\mapsto \mathcal E_{\alpha}(t)$ followed by the limit $\alpha \to 1$, we find 
\be
{\mathcal E}(t) = \lim_{\alpha\to 1} {\mathcal E_{\alpha}}(t) = 2\ln{}{\rm tr}[(M_{\rm l}^\dag M^{\phantom{\dag}}_{\rm r})^{1/2}].
\label{eq:negativityfixed}
\ee
In computing the above limit we used that in this new language the constraint $\braket{{\rm l}}{{\rm r}}=1$ becomes ${\rm tr}[M_{\rm l}^\dag M^{\phantom{\dag}}_{\rm r}]=1$. Eq.~\eqref{eq:negativityfixed} is our \emph{first main result:} it gives a direct connection between the logarithmic negativity at early times and the fixed points of the space transfer matrix. In particular, it allows for an exact evaluation of the former in all cases where the fixed points are known exactly. 

Crucially, even in the cases when the exact form of fixed points is not easily accessible, one can \emph{always} use Eq.~\eqref{eq:negativityfixed} to prove a universal relation between negativity and R\'enyi mutual information. To see this, let us again consider the tripartition in Fig.~\ref{fig:QC} and evaluate the R\'enyi mutual information 
\be
I^{(\alpha)}_{A:B}(t) : = S^{(\alpha)}_{A}(t)+S^{(\alpha)}_{B}(t)-S^{(\alpha)}_{AB}(t), 
\ee
where 
\be
S^{(\alpha)}_{S}(t) = \frac{1}{1-\alpha}\ln{\rm tr}[\rho_S(t)^\alpha],\quad \alpha\in\mathbb R\,,
\ee
is the R\'enyi entropy of the subsystem $S$  at time $t$.   

As shown in Ref.~\cite{klobas2021entanglement}, a space-time duality analysis
along the lines of the one performed above gives 
\begin{align} \label{eq:renyi_entropy}
    \mkern-4mu S^{(n)}_{S}(t) \!=\!\frac{2}{1-n}\ln{}{\rm tr}[(M_{\rm l}^\dag M^{\phantom{\dag}}_{\rm r})^{n}],\quad \mkern-4mu L_S,L_{\bar S} \geq 2v_{\rm max}t.\mkern-4mu
\end{align}
Comparing with Eq.~\eqref{eq:negativityfixed}, we then conclude that in the early-time regime described by Eq.~\eqref{eq:relevantregime}, we have 
\be
2 {\mathcal E}(t) = I^{(1/2)}_{A:B}(t) = S^{(1/2)}_{A}(t) = S^{(1/2)}_{B}(t)\,.
\label{eq:mainresult}
\ee

The universal relation \eqref{eq:mainresult} between negativity and R\'enyi mutual information (and R\'enyi entropies) at early times is our second main result. While for pure states the equivalence trivially follows from the definitions of the quantities involved, for mixed states it is highly non-trivial: negativity measures quantum entanglement while mutual information measures both quantum and classical correlations. Therefore, our results suggests that the correlation built by the dynamics between two
subsystems $A$ and $B$ at early times is of purely quantum nature. 

In fact, using Eq.~\eqref{eq:nega_moment} and Eq.~\eqref{eq:renyi_entropy}, one can straightforwardly generalise Eq.~\eqref{eq:mainresult} to relate the ``moments'' $\mathcal{E}_{2n}(t)$ with the R\'enyi mutual information. In particular, introducing the ratio $\mathcal R_{\alpha}(t)$ as
\be \label{eq:defRatio}
\!
\mathcal{R}_{\alpha}(t) :=
\ln\frac{\tr\left[(\rho(t)^{t_A}_{AB})^{\alpha}\right]}{\tr\left[(\rho(t)_{AB})^{\alpha}\right]}
=\mathcal{E}_{\alpha}(t) -(1-\alpha)S^{(\alpha)}_{AB}(t),\!
\ee
we obtain the following universal relation that holds in the early-time 
regime~\eqref{eq:relevantregime}
\be\label{eq:nega_moment_equlity}
\mathcal{R}_{\alpha}(t)  = \left(1-\frac{\alpha}{2}\right) I^{(\alpha/2)}_{A:B}(t).
\ee
The moments $\mathcal{E}_{\alpha}(t)$, and the ratios $\mathcal{R}_{\alpha}(t)$ are not entanglement monotones but they still return non-trivial information~\cite{Elben2020mixed,zollor_2021_measure_moment,calabrese2012entanglement,Calabrese_2013_nega_extended,Alba_2013_cft_mc,Calabrese_2013_ising,Alba_2014_mc,lu2020detecting,Grover_2020_nega_mc,pollman_2020_mbl,turkeshi2020negativity}. Importantly, in contrast to negativity, this quantities are currently experimentally accessible~\cite{Elben2020mixed,zollor_2021_measure_moment}. Our treatment can also be repeated for other mixed-state entanglement measures --- such as reflected entropy~\cite{dutta2021canonical}, and odd entropy~\cite{tamaoka2019entanglement} --- to show that in the early time regime they are fully specified by the R\'enyi mutual information~\cite{bertini2022earlytime}. This is in agreement with the CFT results of Refs.~\cite{ryu_2020_correlation, kudler2020quasi}.

Remarkably, the relations~\eqref{eq:mainresult}, and~\eqref{eq:nega_moment_equlity} continue to hold for inhomogeneous initial states, and for local gates that are space-time dependent, namely for circuits with any kind of disorder or aperiodic driving. The only notable difference is that for disordered systems in Eq.~\eqref{eq:mainresult} only the first of the equalities holds~\cite{Note1}. This shows that the fundamental equality is the one between negativity and half of the mutual information.  Moreover, our treatment can also be directly applied to initial states that are in the matrix-product-state (MPS) form. For generic homogeneous MPS, one can show that the equalities~\eqref{eq:mainresult}, and~\eqref{eq:nega_moment_equlity} hold up to corrections that are exponentially suppressed in the subsystem sizes~\cite{Note1}. Essentially this is because in these states correlations decay exponentially with the distance.  This reasoning also suggests that one could repeat the same argument also for ground states of critical Hamiltonians. In that case we expect Eqs.~\eqref{eq:mainresult}, and~\eqref{eq:nega_moment_equlity} to hold up to power-law corrections in the size of the subsystems.

One can gain intuition on Eq.~\eqref{eq:mainresult} by considering one-dimensional Clifford circuits~\cite{gottesman1996class} --- quantum circuits mapping Pauli strings to Pauli strings. In these systems, up to unitary operations acting only within given subsystems, the state after a quench from a computational basis state can be decomposed into single-qubit product states, two-qubit Bell pair states with two qubits in two distinct regions, and three-qubit Greenberger-Horne-Zeilinger (GHZ) states with one qubit in each of $A$, $B$, and $C$~\cite{bravyi2006ghz,sang2021entanglement}. As  correlation measures are invariant under unitary transformations, the aforementioned decomposition leads to $\mathcal E = e_{AB} \ln 2$ and $I_{A:B} =( 2e_{AB}+ g_{ABC}) \ln 2$, where $e_{AB}$ and $g_{ABC}$ respectively denote the number of Bell pairs and GHZ states.  Building up non-vanishing connected correlations between any two of the three qubits located in $A$,$B$, and $C$ --- a defining feature of the three-qubit GHZ state --- takes times $t> \text{min}\{L_A,L_B,L_C\}/(2v_{\text{max}})$. It follows that the GHZ-type tripartite correlation cannot exist in the early-time regime~\eqref{eq:relevantregime}, and, consequently, the correlation between any two subsystems solely results from entangled Bell pairs. For product initial states, a similar argument can be used to provide an alternative proof of Eq.~\eqref{eq:mainresult}~\cite{bertini2022earlytime}.

Finally, we stress that the restriction to quantum circuits, i.e., to discrete space-time, is not essential for the validity of Eq.~\eqref{eq:mainresult}. Indeed, the only essential ingredient in our derivation is that the two edges of each subsystem are causally disconnected at early times because of the finite speed for the propagation of signals. Therefore we expect Eq.~\eqref{eq:mainresult} to hold for any quantum many-body system where interactions are local enough to allow for a finite maximal speed. This include quantum spin-chains characterised by the Lieb-Robinson bound~\cite{lieb1972finite}, and relativistic quantum field theories. 

In the future, it would be interesting to go beyond the early-time regime --- at least for special classes of quantum circuits --- and understand under which conditions the relation in Eq.~\eqref{eq:mainresult} ceases to hold. This could potentially distinguish different classes of dynamics, as shown in Ref.~\cite{ryu_2020_correlation} in the case of conformal field theory.

\begin{acknowledgments}
    We thank Jonah Kudler-Flam, Toma{\v z} Prosen and, especially, Pasquale Calabrese for very valuable comments on the manuscript and stimulating discussions. This work has been supported by the Royal Society through the University
Research Fellowship No.\ 201101 (BB), by the EPSRC under grant EP/S020527/1
(KK), and Perimeter Institute for Theoretical Physics (TCL). Research at Perimeter Institute is supported in part by the Government of
Canada through the Department of Innovation, Science and Economic Development
and by the Province of Ontario through the Ministry of Colleges and
Universities. BB and KK thank SISSA for hospitality in the early stage of this project.
\end{acknowledgments}
\bibliography{bibliography}
\onecolumngrid
\newcounter{equationSM}
\newcounter{figureSM}
\newcounter{tableSM}
\stepcounter{equationSM}
\setcounter{equation}{0}
\setcounter{figure}{0}
\setcounter{table}{0}
\makeatletter
\renewcommand{\theequation}{\textsc{sm}-\arabic{equation}}
\renewcommand{\thefigure}{\textsc{sm}-\arabic{figure}}
\renewcommand{\thetable}{\textsc{sm}-\arabic{table}}
\begin{center}
{\large{\bf Supplemental Material for\\
Entanglement Negativity and Mutual Information after a Quantum Quench:\\
    Exact Link from Space-Time Duality}}
\end{center}
Here we report some useful information complementing the main text. In particular
\begin{itemize}
    \item[-] In Sec.~\ref{sec:noTI} we extend the results from the main text to the 
        general case with space-time disorder.
    \item[-] In Sec.~\ref{sec:inMPS} we treat the case with the initial states prepared
        in the matrix-product form.
\end{itemize}

\section{Spatial and temporal inhomogeneities}\label{sec:noTI}
Let us relax the assumption of spatial and temporal homogeneity, i.e.\ the unitary
gates $U_{x,t}$ now depend on the space-time point $(x,t)$, and one-site initial states
$\ket{\psi_x}$ are position dependent. Consequently, the space transfer-matrix $\mathbb{T}_x$
acquires a dependence on the position $x$ (hence the subscript) and has to be defined as,
\be
\mathbb{T}_{x}=\left(
\tilde{\mathbb{U}}_x \otimes \tilde{\mathbb{U}}_x^{\ast}
\right)\cdot\mathbb{O},\qquad
\tilde{\mathbb{U}}_x=
\tilde{\eta}_0\left(\ketbra*{\psi_{x}}{\psi_{x+\frac{1}{2}}}\right)
\mkern-8mu\smashoperator[r]{\prod_{\tau\in\mathbb{Z}_t+1\vphantom{\frac{1}{2}}}} 
\tilde{\eta}_\tau(\tilde{U}_{x,\tau})\mkern-6mu
\mkern-4mu\smashoperator[r]{\prod_{\tau\in\mathbb{Z}_t+\frac{1}{2}}}
\tilde{\eta}_{\tau}(\tilde{U}_{x+\frac{1}{2},\tau}).
\ee
Here $\tilde{U}_{x,t}$ is obtained from $U_{x,t}$ via spacetime duality transformation
(as described in the main text).
Due to the nonhomogeneity it does not make sense to consider powers of the transfer matrix,
and the relevant object becomes a product of consecutive transfer matrices
$\mathbb{T}_x\mathbb{T}_{x+1}\cdots\mathbb{T}_{y-1}$. For $y\geq x+2t$, the unitarity of time-evolution gives us
a relation analogous to Eq.~\eqref{eq:fixedpointrelation}, i.e., 
\be\label{eq:inhomProject}
\begin{aligned}
\mathbb{T}_{x}\mathbb{T}_{x+1}\cdots \mathbb{T}_{y-1}&=
\begin{tikzpicture}
    [baseline={([yshift=-0.6ex]current bounding box.center)},scale=0.45]
    \def\X{18}
    \def\Y{8}
    \begin{scope}[shift={(0.2,0.1)}]
        \tgridLine{0.75}{0.25}{0.25}{0.75}
        \foreach \t in {2,4,...,\Y}{
            \tgridLine{1}{\t}{0.25}{\t-0.75};
            \tgridLine{\X}{\t-1}{\X+0.75}{\t-0.25};
            \tgridLine{1}{\t}{0.25}{\t+0.75}
            \tgridLine{\X}{\t-1}{\X+0.75}{\t-1.75};
        }
        \tgridLine{\X+0.25}{\Y+0.75}{\X+0.75}{\Y+0.25};
        \begin{scope}
            \clip (1,1) rectangle (\X,\Y);
            \foreach \t in {1,3,...,\Y}{\tgridLine{-0.75}{\t-0.75}{\Y+0.75-\t}{\Y+0.75}};
            \foreach \x in {2,4,...,\X}{\tgridLine{\x-0.75}{0.25}{\X+0.75}{\X-\x+1.75}};
            \foreach \t in {-1,1,...,\Y}{\tgridLine{\X+0.75}{\t+1.25}{\t+1}{\X+1}};
            \foreach \x in {1,3,...,\X}{\tgridLine{\x-0.25}{0.25}{0.25}{\x-0.25}};
        \end{scope}
        \foreach\x in {3,5,...,\X}{\tgridLine{\x}{\Y}{\x-0.8}{\Y+0.8};}
        \foreach\x in {1,3,...,\X}{\tgridLine{\x}{\Y}{\x+0.8}{\Y+0.8};}
        \foreach\x in {2,4,...,\X}{\tgridLine{\x}{1}{\x-0.75}{1-0.75};}
        \foreach\x in {4,6,...,\X}{\tgridLine{\x-2}{1}{\x-2+0.75}{1-0.75};}

        \foreach \t in {1,3,...,\Y}{
            \foreach \x in {1,3,...,\X}{\prop{\x}{\t+1}{colUc}};
            \foreach \x in {2,4,...,\X}{\prop{\x}{\t}{colUc}};
        }
        \foreach \x in {1,3,...,\X}{
            \Ips{(\x-0.25)}{0.25}{colIstC}
            \Ips{(\x+0.25)}{0.25}{colIstC}
        }
    \end{scope}
    \tgridLine{0.75}{0.25}{0.25}{0.75}
    \foreach \t in {2,4,...,\Y}{
        \tgridLine{1}{\t}{0.25}{\t-0.75};
        \tgridLine{\X}{\t-1}{\X+0.75}{\t-0.25};
        \tgridLine{1}{\t}{0.25}{\t+0.75}
        \tgridLine{\X}{\t-1}{\X+0.75}{\t-1.75};
    }
    \tgridLine{\X+0.25}{\Y+0.75}{\X+0.75}{\Y+0.25};
    \begin{scope}
        \clip (1,1) rectangle (\X,\Y);
        \foreach \t in {1,3,...,\Y}{\tgridLine{-0.75}{\t-0.75}{\Y+0.75-\t}{\Y+0.75}};
        \foreach \x in {2,4,...,\X}{\tgridLine{\x-0.75}{0.25}{\X+0.75}{\X-\x+1.75}};
        \foreach \t in {-1,1,...,\Y}{\tgridLine{\X+0.75}{\t+1.25}{\t+1}{\X+1}};
        \foreach \x in {1,3,...,\X}{\tgridLine{\x-0.25}{0.25}{0.25}{\x-0.25}};
    \end{scope}
    \foreach\x in {3,5,...,\X}{\tgridLine{\x}{\Y}{\x-0.8}{\Y+0.8};}
    \foreach\x in {1,3,...,\X}{\tgridLine{\x}{\Y}{\x+0.8}{\Y+0.8};}
    \foreach\x in {2,4,...,\X}{\tgridLine{\x}{1}{\x-0.75}{1-0.75};}
    \foreach\x in {4,6,...,\X}{\tgridLine{\x-2}{1}{\x-2+0.75}{1-0.75};}

    \foreach \t in {1,3,...,\Y}{
        \foreach \x in {1,3,...,\X}{\prop{\x}{\t+1}{colU}};
        \foreach \x in {2,4,...,\X}{\prop{\x}{\t}{colU}};
    }
    \foreach \x in {1,3,...,\X}{
            \Ips{(\x-0.25)}{0.25}{colIst}
            \Ips{(\x+0.25)}{0.25}{colIst}
    }
    \foreach \x in {1,3,...,\X}{
        \draw[semithick,colLines,rounded corners=0.5] (\x+0.75,\Y+0.75) -- (\x+1,\Y+1) -- 
        (\x+0.85+0.2,\Y+0.85+0.1) -- (\x+0.75+0.2,\Y+0.75+0.1);
    }
    \foreach \x in {1,3,...,\X}{
        \draw[semithick,colLines,rounded corners=1.5] (\x+2-0.75,\Y+0.75) -- (\x+2-0.9,\Y+0.9) -- 
        (\x+2-0.95+0.2,\Y+0.95+0.1) -- (\x+2-0.75+0.2,\Y+0.75+0.1);
    }
    \node at (1,9) {\scalebox{0.65}{$x$}};
    \node at (2,8) {\scalebox{0.65}{$x\mkern-4mu+\mkern-4mu\frac{1}{2}\mkern-6mu$}};
    \node at (3,9) {\scalebox{0.65}{$x\mkern-4mu+\mkern-4mu 1\mkern-6mu$}};
    \node at (4,8) {\scalebox{0.65}{$x\mkern-4mu+\mkern-4mu\frac{3}{2}\mkern-6mu$}};
    \node at (5,9) {\scalebox{0.65}{$\cdots\mkern-6mu$}};
    \node at (15,9) {\scalebox{0.65}{$\cdots\mkern-6mu$}};
    \node at (16,8) {\scalebox{0.65}{$y\mkern-4mu-\mkern-4mu\frac{3}{2}\mkern-6mu$}};
    \node at (17,9) {\scalebox{0.65}{$y\mkern-4mu-\mkern-4mu 1\mkern-6mu$}};
    \node at (18,8) {\scalebox{0.65}{$y\mkern-4mu-\mkern-4mu\frac{1}{2}\mkern-6mu$}};
\end{tikzpicture}\\
&=
\begin{tikzpicture}
    [baseline={([yshift=-0.6ex]current bounding box.center)},scale=0.45]
    \def\X{18}
    \def\Y{8}
    \begin{scope}[shift={(0.2,0.1)}]
        \foreach \t in {0,2,...,\Y}{\tgridLine{0.25}{\t+0.75}{\t+0.75}{0.25}};
        \foreach \t in {0,2,...,\Y}{\tgridLine{\X+0.25-\t*0.5}{\Y+0.75-\t*0.5}{\X+0.75}{\Y+0.25-\t};}
        \foreach \x in {2,4,6}{\tgridLine{\X-\x+0.75}{0.25}{\X-\x-4+0.5*\x+0.25}{5-0.5*\x-0.25}}
        \foreach \t in {1,3,...,\Y}{\tgridLine{\X+0.75}{\Y-\t+0.75}{\X-\Y+\t+0.25}{0.25}};
        \foreach \t in {2,4,...,\Y}{\tgridLine{0.25}{\t-0.75}{6-0.5*\t-0.25}{5+0.5*\t-0.25}};
        \foreach \x in {2,4,...,8}{\tgridLine{\x-0.75}{0.25}{5+0.5*\x-0.25}{6-0.5*\x-0.25}}
        \tgridLine{9.5}{0.25}{9.5}{1};
        \tgridLine{10.5}{0.25}{10.5}{1};
        \prop{1}{2}{colUc};
        \prop{1}{4}{colUc};
        \prop{1}{6}{colUc};
        \prop{1}{8}{colUc};
        \prop{2}{1}{colUc};
        \prop{2}{3}{colUc};
        \prop{2}{5}{colUc};
        \prop{2}{7}{colUc};
        \prop{3}{2}{colUc};
        \prop{3}{4}{colUc};
        \prop{3}{6}{colUc};
        \prop{4}{1}{colUc};
        \prop{4}{3}{colUc};
        \prop{4}{5}{colUc};
        \prop{5}{2}{colUc};
        \prop{5}{4}{colUc};
        \prop{6}{1}{colUc};
        \prop{6}{3}{colUc};
        \prop{7}{2}{colUc};
        \prop{8}{1}{colUc};

        \prop{18}{1}{colUc};
        \prop{18}{3}{colUc};
        \prop{18}{5}{colUc};
        \prop{18}{7}{colUc};
        \prop{17}{2}{colUc};
        \prop{17}{4}{colUc};
        \prop{17}{6}{colUc};
        \prop{16}{1}{colUc};
        \prop{16}{3}{colUc};
        \prop{16}{5}{colUc};
        \prop{15}{2}{colUc};
        \prop{15}{4}{colUc};
        \prop{14}{1}{colUc};
        \prop{14}{3}{colUc};
        \prop{13}{2}{colUc};
        \prop{12}{1}{colUc};

        \foreach \x in {1,3,5,7,9,13,15,17}{\Ips{(\x-0.25)}{0.25}{colIstC}};
        \foreach \x in {1,3,5,7,11,13,15,17}{\Ips{(\x+0.25)}{0.25}{colIstC}};
        \Ips{9.5}{0.25}{colIstC};
        \Ips{10.5}{0.25}{colIstC};
    \end{scope}
    \foreach \t in {0,2,...,\Y}{\tgridLine{0.25}{\t+0.75}{\t+0.75}{0.25}};
    \foreach \t in {0,2,...,\Y}{\tgridLine{\X+0.25-\t*0.5}{\Y+0.75-\t*0.5}{\X+0.75}{\Y+0.25-\t};}
    \foreach \x in {2,4,6}{\tgridLine{\X-\x+0.75}{0.25}{\X-\x-4+0.5*\x+0.25}{5-0.5*\x-0.25}}
    \foreach \t in {1,3,...,\Y}{\tgridLine{\X+0.75}{\Y-\t+0.75}{\X-\Y+\t+0.25}{0.25}};
    \foreach \t in {2,4,...,\Y}{\tgridLine{0.25}{\t-0.75}{6-0.5*\t-0.25}{5+0.5*\t-0.25}};
    \foreach \x in {2,4,...,8}{\tgridLine{\x-0.75}{0.25}{5+0.5*\x-0.25}{6-0.5*\x-0.25}}

    \tgridLine{9.5}{0.25}{9.5}{1};
    \tgridLine{10.5}{0.25}{10.5}{1};

    \prop{1}{2}{colU};
    \prop{1}{4}{colU};
    \prop{1}{6}{colU};
    \prop{1}{8}{colU};
    \prop{2}{1}{colU};
    \prop{2}{3}{colU};
    \prop{2}{5}{colU};
    \prop{2}{7}{colU};
    \prop{3}{2}{colU};
    \prop{3}{4}{colU};
    \prop{3}{6}{colU};
    \prop{4}{1}{colU};
    \prop{4}{3}{colU};
    \prop{4}{5}{colU};
    \prop{5}{2}{colU};
    \prop{5}{4}{colU};
    \prop{6}{1}{colU};
    \prop{6}{3}{colU};
    \prop{7}{2}{colU};
    \prop{8}{1}{colU};

    \prop{18}{1}{colU};
    \prop{18}{3}{colU};
    \prop{18}{5}{colU};
    \prop{18}{7}{colU};
    \prop{17}{2}{colU};
    \prop{17}{4}{colU};
    \prop{17}{6}{colU};
    \prop{16}{1}{colU};
    \prop{16}{3}{colU};
    \prop{16}{5}{colU};
    \prop{15}{2}{colU};
    \prop{15}{4}{colU};
    \prop{14}{1}{colU};
    \prop{14}{3}{colU};
    \prop{13}{2}{colU};
    \prop{12}{1}{colU};

    \foreach \x in {1,3,5,7,9,13,15,17}{\Ips{(\x-0.25)}{0.25}{colIst}};
    \foreach \x in {1,3,5,7,11,13,15,17}{\Ips{(\x+0.25)}{0.25}{colIst}};
    \Ips{9.5}{0.25}{colIst};
    \Ips{10.5}{0.25}{colIst};

    \foreach \x in {1,...,8}{
        \draw[semithick,colLines,rounded corners=0.5] (\x+0.75,\Y+1.75-\x) -- (\x+1,\Y+2-\x) -- (\x+0.85+0.2,\Y+1.85+0.1-\x) -- (\x+0.75+0.2,\Y+1.75+0.1-\x);
    }
    \foreach \x in {0,...,7}{
        \draw[semithick,colLines,rounded corners=1.5] (1+\X-\x-0.75,\Y-\x+0.75) -- (1+\X-\x-0.9,\Y-\x+0.9) -- (1+\X-\x-0.95+0.2,\Y-\x+0.95+0.1) -- (1+\X-\x-0.75+0.2,\Y-\x+0.75+0.1);
    }
    \foreach \x in {0,1}{
        \draw[semithick,colLines,rounded corners=1.5] (\x+9.5,1) -- (\x+9.5,1.3) -- (\x+9.5+0.2,1.25+0.1) -- (\x+9.5+0.2,1+0.1);
    }

    \draw[very thick,gray,rounded corners=2] (0.55,-0.125) -- (0.55,\Y+1.125) -- (2.25,\Y+1.125)
    -- (9.25,2.125) -- (9.25,-0.125) -- cycle;
    \draw[very thick,gray,rounded corners=2] (\X+0.625,-0.125) -- (\X+0.625,\Y+1.125)
    -- (\X,\Y+1.125) -- (\X-7,2.125) -- (\X-7,-0.125) -- cycle;
    \node at (1,8.875) {\scalebox{0.65}{$x$}};

    \node at (2,8) {\scalebox{0.65}{$x\mkern-4mu+\mkern-4mu\frac{1}{2}\mkern-6mu$}};
    \node at (3,7) {\scalebox{0.65}{$x\mkern-4mu+\mkern-4mu 1\mkern-6mu$}};
    \node at (4,6) {\scalebox{0.65}{$x\mkern-4mu+\mkern-4mu\frac{3}{2}\mkern-6mu$}};
    \node at (5,5) {\scalebox{0.65}{$\ddots$}};
    \node at (17,7) {\scalebox{0.65}{$y\mkern-4mu-\mkern-4mu 1\mkern-6mu$}};
    \node at (18,8) {\scalebox{0.65}{$y\mkern-4mu-\mkern-4mu\frac{1}{2}\mkern-6mu$}};
    \node at (6.5,6.5) {$\ket*{r^{(x)}}$};
    \node at (13.75,6.5) {$\bra*{l^{(y)}}$};
\end{tikzpicture}\\
&=\ket*{r^{(x)}}
\left(\smashoperator[r]{\prod_{r=x+t+1}^{y-t}}
\braket*{\psi_{r-\frac{1}{2}}}\braket*{\psi_{r}}
\right)
\bra*{l^{(y)}}=\ketbra*{r^{(x)}}{l^{(y)}},
\qquad y-x\ge 2t.
\end{aligned}
\ee
Here the last equality follows from normalization of the initial state, 
\be
\braket{\psi_{x}}=1,
\ee
while the second equality from the repeated use of the unitarity of local gates,
\be
U_{x,t}U_{x,t}^{\dagger}=\1 \Leftrightarrow
\begin{tikzpicture}
    [baseline={([yshift=-0.6ex]current bounding box.center)},scale=0.45]
    \begin{scope}[shift={(0.2,0.1)}]
        \tgridLine{-0.75}{-0.75}{0.75}{0.75};
        \tgridLine{-0.75}{0.75}{0.75}{-0.75};
        \prop{0}{0}{colUc};
    \end{scope}
    \tgridLine{-0.75}{-0.75}{0.75}{0.75};
    \tgridLine{-0.75}{0.75}{0.75}{-0.75};
    \prop{0}{0}{colU};
    \draw[semithick,colLines,rounded corners=0.5] (0.75,0.75) -- (1,1) -- (0.85+0.2,0.85+0.1) -- (0.75+0.2,0.75+0.1);
    \draw[semithick,colLines,rounded corners=1.5] (-0.75,0.75) -- (-0.9,0.9) -- (-0.95+0.2,0.95+0.1) -- (-0.75+0.2,0.75+0.1);
\end{tikzpicture}=
\begin{tikzpicture}
    [baseline={([yshift=-0.6ex]current bounding box.center)},scale=0.45]
    \begin{scope}[shift={(0.2,0.1)}]
        \tgridLine{-0.25}{0.75}{-1+0.25}{0.25};
        \tgridLine{0.25}{0.75}{1-0.25}{0.25};
    \end{scope}
    \tgridLine{-0.25}{0.75}{-1+0.25}{0.25};
    \tgridLine{0.25}{0.75}{1-0.25}{0.25};
    \draw[semithick,colLines,rounded corners=0.5] (-0.25,0.75) -- (0,1) -- (-0.15+0.2,0.85+0.1) -- (-0.25+0.2,0.75+0.1);
    \draw[semithick,colLines,rounded corners=1.5] (0.25,0.75) -- (0.1,0.9) -- (0.05+0.2,0.95+0.1) -- (0.25+0.2,0.75+0.1);
\end{tikzpicture}.
\ee
Note that the factorization relation~\eqref{eq:inhomProject} reduces precisely to Eq.~\eqref{eq:fixedpointrelation} when translational invariance is assumed.

Introducing now the shorthand notation
\be
\mathbb{T}^{[x,y]}=
\mathbb{T}_x\mathbb{T}_{x+1}\mathbb{T}_{x+2}\cdots \mathbb{T}_{y-1},\qquad
\mathbb{T}^{[x,y]}_{2n}=\left.\mathbb{T}^{[x,y]}\right.^{\otimes n},
\ee
and denoting the positions of interfaces between the subsystems by
$x_{AB}$, $x_{BC}$ and $x_{CA}$,
the analogue of Eq.~\eqref{eq:replicasrewritten} can be rewritten as
\be
\mathcal{E}_{2n}=\ln \tr[\mathbb{P}_{\pi_1}^{\dagger}
    \mathbb{T}^{[x_{CA},x_{AB}]}_{2n} \mathbb{P}_{\pi_1}
    \mathbb{P}_{\pi_2}^{\dagger} \mathbb{T}^{[x_{AB},x_{BC}]}_{2n}
    \mathbb{P}_{\pi_2} \mathbb{T}^{[x_{BC},x_{CA}]}_{2n}].
\ee
Combining this with~\eqref{eq:inhomProject} we see that whenever 
\be
L_{A},L_{B},L_{C}\ge2t,
\ee
we obtain the following expression
\be
\begin{aligned}
    \mathcal{E}_{2n}(t)&=
\ln[
    \mel*{l_{2n}^{(x_{CA})}}{\mathbb{P}_{\pi_1}^{\dagger}}{r_{2n}^{(x_{CA})}}
    \mel*{l_{2n}^{(x_{AB})}}{\mathbb{P}_{\pi_1}\mathbb{P}_{\pi_2}^{\dagger}}{r_{2n}^{(x_{AB})}}
    \mel*{l_{2n}^{(x_{BC})}}{\mathbb{P}_{\pi_2}}{r_{2n}^{(x_{BC})}}
]\\
    &=\ln\big(
    \tr [(M_{\mathrm{l}}^{(x_{CA})\dagger}M_{\mathrm{r}}^{(x_{CA})})^{2n}]
    \tr [(M_{\mathrm{l}}^{(x_{BC})\dagger}M_{\mathrm{r}}^{(x_{BC})})^{2n}]
    \left(\tr [(M_{\mathrm{l}}^{(x_{AB})\dagger}M_{\mathrm{r}}^{(x_{AB})})^{n}]\right)^2
\big),
\end{aligned}
\ee
where $\bra*{l_{2n}^{(x)}}$, $\ket*{r_{2n}^{(x)}}$, $M_{\mathrm{l}}^{(x)}$, and
$M_{\mathrm{r}}^{(x)}$ are defined analogously to the position independent
quantities in the main text (see Eq.~\eqref{eq:vtomapping} and text after
Eq.~\eqref{eq:replicassimple}).  The normalization condition
$\tr[M_{\mathrm{l}}^{(x)\dagger}M_{\mathrm{r}}^{(y)}]=\braket{l^{(x)}}{r^{(y)}}=1$
for arbitrary $x$, $y$ implies that in the expression for negativity
$\mathcal{E}(t)=\lim_{\alpha\to 1}\mathcal{E}_{\alpha}(t)$ is analogous to the
homogeneous expression~\eqref{eq:negativityfixed}, where the matrices
$M_{\mathrm{l/r}}^{(x)}$ are evaluated at position $x=x_{AB}$,
\be
\mathcal{E}(t)=2\ln
    \tr [(M_{\mathrm{l}}^{(x_{AB})\dagger}M_{\mathrm{r}}^{(x_{AB})})^{\frac{1}{2}}].
\ee
Similarly, one can evaluate R\'enyi entropies of all three subsystems
\be
\begin{aligned}
    S_{A}^{(n)}&=\frac{1}{1-n}
\left( \ln \tr[(M_{\mathrm{l}}^{(x_{CA})\dagger}M_{\mathrm{r}}^{(x_{CA})})^n]
+\ln \tr[(M_{\mathrm{l}}^{(x_{AB})\dagger}M_{\mathrm{r}}^{(x_{AB})})^n]\right),\\
    S_{B}^{(n)}&=\frac{1}{1-n}
\left( \ln \tr[(M_{\mathrm{l}}^{(x_{AB})\dagger}M_{\mathrm{r}}^{(x_{AB})})^n]
+\ln \tr[(M_{\mathrm{l}}^{(x_{BC})\dagger}M_{\mathrm{r}}^{(x_{BC})})^n]\right),\\
    S_{AB}^{(n)}&=\frac{1}{1-n}
\left( \ln \tr[(M_{\mathrm{l}}^{(x_{CA})\dagger}M_{\mathrm{r}}^{(x_{CA})})^n]
+\ln \tr[(M_{\mathrm{l}}^{(x_{BC})\dagger}M_{\mathrm{r}}^{(x_{BC})})^n]\right)
    \end{aligned},
\ee
which implies the expression for the mutual information that depends only on the 
matrices $M_{\mathrm{l/r}}^{(x)}$ evaluated at the edge between subsystems 
$A$ and $B$
\be
I_{A:B}^{(n)}(t)=\frac{2}{1-n}
\ln \tr[(M_{\mathrm{l}}^{(x_{AB})\dagger}M_{\mathrm{r}}^{(x_{AB})})^n].
\ee
This finally gives us the same connection between the R\'enyi-$\frac{1}{2}$ mutual
information and negativity that holds also in the homogeneous case,
\be
\mathcal{E}(t)=\frac{1}{2}I_{A:B}^{(\frac{1}{2})}.
\ee
However, there is an important difference --- in the inhomogeneous setting the relation
holds \emph{only} between the negativity and mutual information, as R\'enyi entropies 
of subsystems are now position dependent and are no longer the same.

Similarly, the relation between the moments $\varepsilon_{\alpha}(t)$ and the
ratios $\mathcal{R}_{\alpha}(t)$ (cf.\ Eq.~\eqref{eq:defRatio}) can be shown to
hold also in the inhomogeneous case,
\be
\mathcal{R}_{\alpha}(t)=
    \mathcal{E}_{\alpha}(t)-(1-\alpha)S_{AB}^{(\alpha)}(t)=
    2\ln\tr [(M_{\mathrm{l}}^{(x_{AB})\dagger}M_{\mathrm{r}}^{(x_{AB})})^{\frac{\alpha}{2}}]
    =\left(1-\frac{\alpha}{2}\right)I_{A:B}^{(\frac{\alpha}{2})}.
\ee

\section{Matrix-product initial state}\label{sec:inMPS}
\begin{figure}
    \includegraphics[width=0.5\columnwidth]{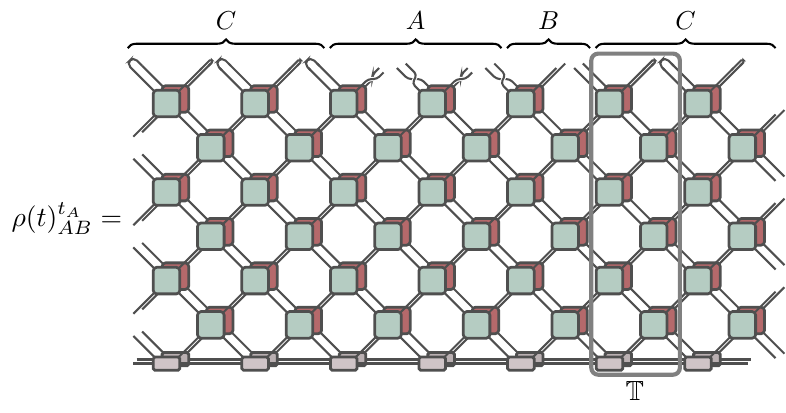}
    \caption{\label{fig:foldedCircuitsm} Folded representation of $\rho(t)_{AB}^{t_A}$
    starting from an initial state in the MPS form. The diagram is analogous to the 
    one in Fig.~\ref{fig:foldedCircuit}, but the initial state is now no longer a product
    state. As a consequence the space transfer matrix now acts on the Hilbert space that also
    includes two copies of the auxiliary space of the initial MPS.
    }
\end{figure}

Let us now consider a two-site translationally invariant initial state in the following form
\be
\ket{\Psi_0}=
\smashoperator{\sum_{s_1,\ldots,s_{2L}=1}^{d}}
\tr\left[W^{s_1s_2} W^{s_3s_4}\cdots W^{s_{2L-1}s_{2L}}\right]
\ket{s_1}\otimes\cdots\otimes\ket{s_{2L}},
\label{eq:psi0}
\ee  
where $W^{sr}$ are $\chi\times\chi$ matrices ($\chi$ is an arbitrary positive
integer), with matrix elements $W^{sr}_{\mu\nu}$, and $\{\ket{s}\}_{s=1}^{d}$
is the canonical orthonormal basis of $\mathbb C^d$. Namely, $\ket{\Psi_0}$ is
a two-site matrix-product state (MPS) with \emph{bond dimension} $\chi$.
The space transfer-matrix $\mathbb{T}_x\in\rm{End}(\mathcal{H}_t^{\otimes 2})$
now acts on a larger Hilbert space with vectors describing a state of $2t+1$ qudits
and the auxiliary space of the MPS,
\be
\mathcal{H}_t=\mathbb{C}^{\chi d^{2t+1}}.
\ee
In particular, $\mathbb{T}_x$ takes the following form (see Fig.~\ref{fig:foldedCircuitsm}
for a diagrammatic illustration)
\be
\mathbb{T}_{x}=\left(
\tilde{\mathbb{U}}_x \otimes \tilde{\mathbb{U}}_x^{\ast}
\right)\cdot\mathbb{O},\qquad
\tilde{\mathbb{U}}_x=
\tilde{\eta}_0\left(\tilde{W}\right)
\mkern-8mu\smashoperator[r]{\prod_{\tau\in\mathbb{Z}_t+1\vphantom{\frac{1}{2}}}} 
\tilde{\eta}_\tau(\tilde{U}_{x,\tau})\mkern-6mu
\mkern-4mu\smashoperator[r]{\prod_{\tau\in\mathbb{Z}_t+\frac{1}{2}}}
\tilde{\eta}_{\tau}(\tilde{U}_{x+\frac{1}{2},\tau}),
\ee
where $\tilde{W}$ is obtained from $W$ by applying the spacetime-duality transformation.

The factorization property does not hold immediately, but depends on the properties of the
MPS. Indeed, after taking into account the unitarity of local gates the product of 
transfer matrices is expressed as
\be\label{eq:mpsProject}
\mathbb{T}^{[x,y]}=
\begin{tikzpicture}
    [baseline={([yshift=-0.6ex]current bounding box.center)},scale=0.45]
    \def\X{24}
    \def\Y{8}
    \begin{scope}[shift={(0.2,0.1)}]
    \foreach \t in {0,2,...,\Y}{\tgridLine{0.25}{\t+0.75}{\t+0.875}{0.125}};
    \foreach \t in {0,2,...,\Y}{\tgridLine{\X+0.25-\t*0.5}{\Y+0.75-\t*0.5}{\X+0.75}{\Y+0.25-\t};}
    \foreach \x in {2,4,6}{\tgridLine{\X-\x+0.75}{0.25}{\X-\x-4+0.5*\x+0.25}{5-0.5*\x-0.25}}
    \foreach \t in {1,3,...,\Y}{\tgridLine{\X+0.75}{\Y-\t+0.75}{\X-\Y+\t+0.125}{0.125}};
    \foreach \t in {2,4,...,\Y}{\tgridLine{0.25}{\t-0.75}{6-0.5*\t-0.25}{5+0.5*\t-0.25}};
    \foreach \x in {2,4,...,8}{\tgridLine{\x-0.875}{0.125}{5+0.5*\x-0.25}{6-0.5*\x-0.25}}

        \foreach \x in {9,11,13,15}{\tgridLine{\x+0.25}{0.125}{\x+0.25}{1}};
        \foreach \x in {11,13,15,17}{\tgridLine{\x-0.25}{0.125}{\x-0.25}{1}};
        \prop{1}{2}{colUc};
        \prop{1}{4}{colUc};
        \prop{1}{6}{colUc};
        \prop{1}{8}{colUc};
        \prop{2}{1}{colUc};
        \prop{2}{3}{colUc};
        \prop{2}{5}{colUc};
        \prop{2}{7}{colUc};
        \prop{3}{2}{colUc};
        \prop{3}{4}{colUc};
        \prop{3}{6}{colUc};
        \prop{4}{1}{colUc};
        \prop{4}{3}{colUc};
        \prop{4}{5}{colUc};
        \prop{5}{2}{colUc};
        \prop{5}{4}{colUc};
        \prop{6}{1}{colUc};
        \prop{6}{3}{colUc};
        \prop{7}{2}{colUc};
        \prop{8}{1}{colUc};

        \prop{\X}{1}{colUc};
        \prop{\X}{3}{colUc};
        \prop{\X}{5}{colUc};
        \prop{\X}{7}{colUc};
        \prop{\X-1}{2}{colUc};
        \prop{\X-1}{4}{colUc};
        \prop{\X-1}{6}{colUc};
        \prop{\X-2}{1}{colUc};
        \prop{\X-2}{3}{colUc};
        \prop{\X-2}{5}{colUc};
        \prop{\X-3}{2}{colUc};
        \prop{\X-3}{4}{colUc};
        \prop{\X-4}{1}{colUc};
        \prop{\X-4}{3}{colUc};
        \prop{\X-5}{2}{colUc};
        \prop{\X-6}{1}{colUc};

        \draw [thick,colLines] (0.125,0.125) -- (\X,0.125);
        \foreach \x in {1,3,...,\X}{\ImpsM{(\x)}{0.125}{colIstC}};

    \end{scope}
    \foreach \t in {0,2,...,\Y}{\tgridLine{0.25}{\t+0.75}{\t+0.875}{0.125}};
    \foreach \t in {0,2,...,\Y}{\tgridLine{\X+0.25-\t*0.5}{\Y+0.75-\t*0.5}{\X+0.75}{\Y+0.25-\t};}
    \foreach \x in {2,4,6}{\tgridLine{\X-\x+0.75}{0.25}{\X-\x-4+0.5*\x+0.25}{5-0.5*\x-0.25}}
    \foreach \t in {1,3,...,\Y}{\tgridLine{\X+0.75}{\Y-\t+0.75}{\X-\Y+\t+0.125}{0.125}};
    \foreach \t in {2,4,...,\Y}{\tgridLine{0.25}{\t-0.75}{6-0.5*\t-0.25}{5+0.5*\t-0.25}};
    \foreach \x in {2,4,...,8}{\tgridLine{\x-0.875}{0.125}{5+0.5*\x-0.25}{6-0.5*\x-0.25}}

    \foreach \x in {9,11,13,15}{\tgridLine{\x+0.25}{0.125}{\x+0.25}{1}};
    \foreach \x in {11,13,15,17}{\tgridLine{\x-0.25}{0.125}{\x-0.25}{1}};

    \prop{1}{2}{colU};
    \prop{1}{4}{colU};
    \prop{1}{6}{colU};
    \prop{1}{8}{colU};
    \prop{2}{1}{colU};
    \prop{2}{3}{colU};
    \prop{2}{5}{colU};
    \prop{2}{7}{colU};
    \prop{3}{2}{colU};
    \prop{3}{4}{colU};
    \prop{3}{6}{colU};
    \prop{4}{1}{colU};
    \prop{4}{3}{colU};
    \prop{4}{5}{colU};
    \prop{5}{2}{colU};
    \prop{5}{4}{colU};
    \prop{6}{1}{colU};
    \prop{6}{3}{colU};
    \prop{7}{2}{colU};
    \prop{8}{1}{colU};

    \prop{\X}{1}{colU};
    \prop{\X}{3}{colU};
    \prop{\X}{5}{colU};
    \prop{\X}{7}{colU};
    \prop{\X-1}{2}{colU};
    \prop{\X-1}{4}{colU};
    \prop{\X-1}{6}{colU};
    \prop{\X-2}{1}{colU};
    \prop{\X-2}{3}{colU};
    \prop{\X-2}{5}{colU};
    \prop{\X-3}{2}{colU};
    \prop{\X-3}{4}{colU};
    \prop{\X-4}{1}{colU};
    \prop{\X-4}{3}{colU};
    \prop{\X-5}{2}{colU};
    \prop{\X-6}{1}{colU};

    \draw [thick,colLines] (0.125,0.125) -- (\X,0.125);
    \foreach \x in {1,3,...,\X}{\ImpsM{(\x)}{0.125}{colIst}};

    \foreach \x in {1,...,8}{
        \draw[semithick,colLines,rounded corners=0.5] (\x+0.75,\Y+1.75-\x) -- (\x+1,\Y+2-\x) -- (\x+0.85+0.2,\Y+1.85+0.1-\x) -- (\x+0.75+0.2,\Y+1.75+0.1-\x);
    }
    \foreach \x in {0,...,7}{
        \draw[semithick,colLines,rounded corners=1.5] (1+\X-\x-0.75,\Y-\x+0.75) -- (1+\X-\x-0.9,\Y-\x+0.9) -- (1+\X-\x-0.95+0.2,\Y-\x+0.95+0.1) -- (1+\X-\x-0.75+0.2,\Y-\x+0.75+0.1);
    }
    \foreach \x in {9,11,13,15}{
        \draw[semithick,colLines,rounded corners=1.5] (\x+0.25,1) -- (\x+0.25,1.3) -- (\x+0.25+0.2,1.25+0.1) -- (\x+0.25+0.2,1+0.1);
    }
    \foreach \x in {11,13,15,17}{
        \draw[semithick,colLines,rounded corners=1.5] (\x-0.25,1) -- (\x-0.25,1.3) -- (\x-0.25+0.2,1.25+0.1) -- (\x-0.25+0.2,1+0.1);
    }

    \node at (1,9) {\scalebox{0.65}{$x$}};
    \node at (2,8) {\scalebox{0.65}{$x\mkern-4mu+\mkern-4mu\frac{1}{2}\mkern-6mu$}};
    \node at (3,7) {\scalebox{0.65}{$x\mkern-4mu+\mkern-4mu 1\mkern-6mu$}};
    \node at (4,6) {\scalebox{0.65}{$x\mkern-4mu+\mkern-4mu\frac{3}{2}\mkern-6mu$}};
    \node at (5,5) {\scalebox{0.65}{$\ddots$}};
    \node at (\X-1,7) {\scalebox{0.65}{$y\mkern-4mu-\mkern-4mu 1\mkern-6mu$}};
    \node at (\X,8) {\scalebox{0.65}{$y\mkern-4mu-\mkern-4mu\frac{1}{2}\mkern-6mu$}};

    \draw [semithick,decorate,decoration={brace}](15.5,-0.125)--(10.5,-0.125)
    node [midway,yshift=-8pt] {$\tau^{y-x-2t-1}$};
\end{tikzpicture},
\ee
where we introduced the \emph{MPS transfer matrix} $\tau$ given by the following 
matrix elements
\be
\tau_{\mu_1\nu_1}^{\mu_2\nu_2}=
\smashoperator{\sum_{s_1,s_{2}=1}^{d}} 
W^{s_1s_2}_{\mu_1\mu_2} (W^{s_1s_2}_{\nu_1\nu_2})^*.
\label{eq:MPStm}
\ee
To be able to further simplify the diagram~\eqref{eq:mpsProject} we have to
assume that the maximal eigenvalue of the MPS transfer matrix is unique, which
means that the MPS is \emph{injective}~\cite{cirac2021matrix}.
This is a crucial assumption, as a non-injective MPS might not obey the
relation~\eqref{eq:mainresult} even in the initial state. A simple example is
the GHZ state
\be
\ket{\rm GHZ}=\frac{1}{\sqrt{2}}\left(\ket{00\cdots 0}+\ket{11\cdots 1}\right).
\ee
For this state, the reduced density matrix $\rho_{AB}^{\rm GHZ}$ can be shown to
have zero negativity, while the mutual information takes a finite value
\be
\mathcal{E}(\rho_{AB}^{\rm GHZ})=0,\qquad
I_{A:B}^{(n)}(\rho_{AB}^{\rm GHZ})=\ln 2,
\ee
which violates Eq.~\eqref{eq:mainresult}.

Without the loss of generality we can set the largest eigenvalue of the transfer-matrix of an
injective MPS to be equal to $1$, and denote by $\lambda$ the maximal magnitude of 
subleading eigenvalues. In this case, up to errors of the order
$\lambda^{n}$, the $n$-th power of $\tau$ can be replaced by a projector involving leading eigenvectors 
\be
\tau^{n}=\ketbra{r_{\tau}}{l_{\tau}} + \mathcal{O}(\lambda^{n})
\ee
where $\ket{r_{\tau}}$ and $\bra{l_{\tau}}$ are right and left leading eigenvectors of $\tau$.
This immediately implies that also $\mathbb{T}^{[x,y]}$ can be factorized up to exponentially
small corrections
\be \label{eq:MPSfactor}
\mathbb{T}^{[x,y]}=\ketbra*{r^{(x)}}{l^{(y)}}+\mathcal{O}\left({\lambda^{y-x-2t-1}}\right),\qquad
y-x> 2t,
\ee
where $\ket*{r^{(x)}}$ and $\bra*{l^{(y)}}$ are defined as
\be\label{eq:fixedpointsMPS}
\ket*{r^{(x)}}=
\begin{tikzpicture}
    [baseline={([yshift=-0.6ex]current bounding box.center)},scale=0.45]
    \def\X{24}
    \def\Y{8}

    \begin{scope}[shift={(0.2,0.1)}]
    \foreach \t in {0,2,...,\Y}{\tgridLine{0.25}{\t+0.75}{\t+0.875}{0.125}};
    \foreach \t in {2,4,...,\Y}{\tgridLine{0.25}{\t-0.75}{6-0.5*\t-0.25}{5+0.5*\t-0.25}};
    \foreach \x in {2,4,...,8}{\tgridLine{\x-0.875}{0.125}{5+0.5*\x-0.25}{6-0.5*\x-0.25}}

        \foreach \x in {9}{\tgridLine{\x+0.25}{0.125}{\x+0.25}{1}};
        \prop{1}{2}{colUc};
        \prop{1}{4}{colUc};
        \prop{1}{6}{colUc};
        \prop{1}{8}{colUc};
        \prop{2}{1}{colUc};
        \prop{2}{3}{colUc};
        \prop{2}{5}{colUc};
        \prop{2}{7}{colUc};
        \prop{3}{2}{colUc};
        \prop{3}{4}{colUc};
        \prop{3}{6}{colUc};
        \prop{4}{1}{colUc};
        \prop{4}{3}{colUc};
        \prop{4}{5}{colUc};
        \prop{5}{2}{colUc};
        \prop{5}{4}{colUc};
        \prop{6}{1}{colUc};
        \prop{6}{3}{colUc};
        \prop{7}{2}{colUc};
        \prop{8}{1}{colUc};
        \draw [thick,colLines] (0.125,0.125) -- (10,0.125);
        \foreach \x in {1,3,...,9}{\ImpsM{(\x)}{0.125}{colIstC}};
    \end{scope}
    \foreach \t in {0,2,...,\Y}{\tgridLine{0.25}{\t+0.75}{\t+0.875}{0.125}};
    \foreach \t in {2,4,...,\Y}{\tgridLine{0.25}{\t-0.75}{6-0.5*\t-0.25}{5+0.5*\t-0.25}};
    \foreach \x in {2,4,...,8}{\tgridLine{\x-0.875}{0.125}{5+0.5*\x-0.25}{6-0.5*\x-0.25}}

    \foreach \x in {9}{\tgridLine{\x+0.25}{0.125}{\x+0.25}{1}};

    \prop{1}{2}{colU};
    \prop{1}{4}{colU};
    \prop{1}{6}{colU};
    \prop{1}{8}{colU};
    \prop{2}{1}{colU};
    \prop{2}{3}{colU};
    \prop{2}{5}{colU};
    \prop{2}{7}{colU};
    \prop{3}{2}{colU};
    \prop{3}{4}{colU};
    \prop{3}{6}{colU};
    \prop{4}{1}{colU};
    \prop{4}{3}{colU};
    \prop{4}{5}{colU};
    \prop{5}{2}{colU};
    \prop{5}{4}{colU};
    \prop{6}{1}{colU};
    \prop{6}{3}{colU};
    \prop{7}{2}{colU};
    \prop{8}{1}{colU};

    \draw [thick,colLines] (0.125,0.125) -- (10,0.125);
    \foreach \x in {1,3,...,9}{\ImpsM{(\x)}{0.125}{colIst}};

    \foreach \x in {1,...,8}{
        \draw[semithick,colLines,rounded corners=0.5] (\x+0.75,\Y+1.75-\x) -- (\x+1,\Y+2-\x) -- (\x+0.85+0.2,\Y+1.85+0.1-\x) -- (\x+0.75+0.2,\Y+1.75+0.1-\x);
    }
    \foreach \x in {9}{
        \draw[semithick,colLines,rounded corners=1.5] (\x+0.25,1) -- (\x+0.25,1.3) -- (\x+0.25+0.2,1.25+0.1) -- (\x+0.25+0.2,1+0.1);
    }
    
    \draw[thick,colLines,fill=colIst,rounded corners=0.5] (10-0.1,0.125-0.15-0.05) -- (10+0.3,0.125-0.15+0.15) -- (10+0.3,0.125+0.15+0.15) -- (10-0.1,0.125+0.15-0.05) -- cycle;

    \node at (1,9) {\scalebox{0.65}{$x$}};
    \node at (2,8) {\scalebox{0.65}{$x\mkern-4mu+\mkern-4mu\frac{1}{2}\mkern-6mu$}};
    \node at (3,7) {\scalebox{0.65}{$x\mkern-4mu+\mkern-4mu 1\mkern-6mu$}};
    \node at (4,6) {\scalebox{0.65}{$x\mkern-4mu+\mkern-4mu\frac{3}{2}\mkern-6mu$}};
    \node at (5,5) {\scalebox{0.65}{$\ddots$}};
    \draw [semithick,decorate,decoration={brace}](9.75,0.5)--(10.5,0.5)
    node [midway,yshift=10pt] {$\ \ket*{r_{\tau}}$};
\end{tikzpicture},\qquad
\bra*{l^{(y)}}=
\begin{tikzpicture}
    [baseline={([yshift=-0.6ex]current bounding box.center)},scale=0.45]
    \def\X{24}
    \def\Y{8}
    \draw[thick,colLines,fill=colIst,rounded corners=0.5] (16-0.1,0.125-0.15-0.05) -- (16+0.3,0.125-0.15+0.15) -- (16+0.3,0.125+0.15+0.15) -- (16-0.1,0.125+0.15-0.05) -- cycle;
    \begin{scope}[shift={(0.2,0.1)}]
    \foreach \t in {0,2,...,\Y}{\tgridLine{\X+0.25-\t*0.5}{\Y+0.75-\t*0.5}{\X+0.75}{\Y+0.25-\t};}
    \foreach \x in {2,4,6}{\tgridLine{\X-\x+0.75}{0.25}{\X-\x-4+0.5*\x+0.25}{5-0.5*\x-0.25}}
    \foreach \t in {1,3,...,\Y}{\tgridLine{\X+0.75}{\Y-\t+0.75}{\X-\Y+\t+0.125}{0.125}};
        \foreach \x in {17}{\tgridLine{\x-0.25}{0.125}{\x-0.25}{1}};
        \prop{\X}{1}{colUc};
        \prop{\X}{3}{colUc};
        \prop{\X}{5}{colUc};
        \prop{\X}{7}{colUc};
        \prop{\X-1}{2}{colUc};
        \prop{\X-1}{4}{colUc};
        \prop{\X-1}{6}{colUc};
        \prop{\X-2}{1}{colUc};
        \prop{\X-2}{3}{colUc};
        \prop{\X-2}{5}{colUc};
        \prop{\X-3}{2}{colUc};
        \prop{\X-3}{4}{colUc};
        \prop{\X-4}{1}{colUc};
        \prop{\X-4}{3}{colUc};
        \prop{\X-5}{2}{colUc};
        \prop{\X-6}{1}{colUc};

        \draw [thick,colLines] (16,0.125) -- (\X,0.125);
        \foreach \x in {17,19,...,\X}{\ImpsM{(\x)}{0.125}{colIstC}};

    \end{scope}
    \foreach \t in {0,2,...,\Y}{\tgridLine{\X+0.25-\t*0.5}{\Y+0.75-\t*0.5}{\X+0.75}{\Y+0.25-\t};}
    \foreach \x in {2,4,6}{\tgridLine{\X-\x+0.75}{0.25}{\X-\x-4+0.5*\x+0.25}{5-0.5*\x-0.25}}
    \foreach \t in {1,3,...,\Y}{\tgridLine{\X+0.75}{\Y-\t+0.75}{\X-\Y+\t+0.125}{0.125}};

    \foreach \x in {17}{\tgridLine{\x-0.25}{0.125}{\x-0.25}{1}};

    \prop{\X}{1}{colU};
    \prop{\X}{3}{colU};
    \prop{\X}{5}{colU};
    \prop{\X}{7}{colU};
    \prop{\X-1}{2}{colU};
    \prop{\X-1}{4}{colU};
    \prop{\X-1}{6}{colU};
    \prop{\X-2}{1}{colU};
    \prop{\X-2}{3}{colU};
    \prop{\X-2}{5}{colU};
    \prop{\X-3}{2}{colU};
    \prop{\X-3}{4}{colU};
    \prop{\X-4}{1}{colU};
    \prop{\X-4}{3}{colU};
    \prop{\X-5}{2}{colU};
    \prop{\X-6}{1}{colU};

    \draw [thick,colLines] (16,0.125) -- (\X,0.125);
    \foreach \x in {17,19,...,\X}{\ImpsM{(\x)}{0.125}{colIst}};

    \foreach \x in {0,...,7}{
        \draw[semithick,colLines,rounded corners=1.5] (1+\X-\x-0.75,\Y-\x+0.75) -- (1+\X-\x-0.9,\Y-\x+0.9) -- (1+\X-\x-0.95+0.2,\Y-\x+0.95+0.1) -- (1+\X-\x-0.75+0.2,\Y-\x+0.75+0.1);
    }
    \foreach \x in {17}{
        \draw[semithick,colLines,rounded corners=1.5] (\x-0.25,1) -- (\x-0.25,1.3) -- (\x-0.25+0.2,1.25+0.1) -- (\x-0.25+0.2,1+0.1);
    }

    \node at (\X-1,7) {\scalebox{0.65}{$y\mkern-4mu-\mkern-4mu 1\mkern-6mu$}};
    \node at (\X,8) {\scalebox{0.65}{$y\mkern-4mu-\mkern-4mu\frac{1}{2}\mkern-6mu$}};
    \draw [semithick,decorate,decoration={brace}](15.75,0.5)--(16.5,0.5)
    node [midway,yshift=10pt] {$\bra*{l_{\tau}}\ $};
\end{tikzpicture}.
\ee

Having established the factorization property~\eqref{eq:MPSfactor}, we can now repeat the same
reasoning as in Sec.~\ref{sec:noTI}, and obtain the relation
\be
\mathcal{E}(t)\simeq\frac{1}{2} I_{A:B}^{(\frac{1}{2})}(t),\qquad
\mathcal{R}_{\alpha}(t)=\left(1-\frac{\alpha}{2}\right) I_{A:B}^{(\frac{\alpha}{2})}(t),
\ee
where $\simeq$ indicates equality up to corrections of order
\be
\mathcal{O}(\lambda^{L_{m}-2t-1}),\qquad
L_{m}=\min\{L_{A},L_{B},L_{C}\}.
\ee
The initial MPS therefore implies that the relations are valid up to corrections exponentially
small in the subsystem sizes.

Here we assumed only the initial state to be translationally invariant, while
we allowed for the disorder in the local gates. In principle this assumption can
be relaxed, and the MPS matrices can depend on the position $x$, $W\to W_{x}$ and
$\tau\to \tau_x$. In this case, we still require a relation analogous to
Eq.~\eqref{eq:MPSfactor}; namely, there exists $\lambda<1$ so that
\be
\tau_x \tau_{x+1}\cdots \tau_{x+{n-1}} = 
\ketbra*{r_{\tau}^{(x)}}{l_{\tau}^{(x+n)}} + \mathcal{O}(\lambda^{n}),
\ee
for some vectors $\ket*{r_{\tau}^{(x)}}$, and $\bra*{l_{\tau}^{(x+n)}}$. If the MPS
is translationally invariant, this condition immediately holds by requiring injectiveness,
while generically the physical assumptions behind the MPS are less direct.




\end{document}